%% file: main.tex
\documentclass[preprint,12pt,onecolumn,draftclsnofoot,]{IEEEtran}
\IEEEoverridecommandlockouts
\usepackage{mathtools}
\usepackage{amssymb,amsmath}
\usepackage{ifpdf}
\usepackage{url}
\usepackage{graphicx}
\usepackage{caption}
\usepackage{subcaption}
\usepackage{cite}
\usepackage{etoolbox}
\usepackage{multirow}
\usepackage{tabu}
\usepackage{wrapfig}
\usepackage{breqn}
\usepackage{array}
\usepackage{booktabs}

\usepackage{hyperref}

\DeclareUnicodeCharacter{2212}{\ensuremath{-}}

\newcolumntype{L}[1]{>{\raggedright\let\newline\\\arraybackslash\hspace{0pt}}m{#1}}
\newcolumntype{C}[1]{>{\centering\let\newline\\\arraybackslash\hspace{0pt}}m{#1}}
\newcolumntype{R}[1]{>{\raggedleft\let\newline\\\arraybackslash\hspace{0pt}}m{#1}}

\begin{document}

\title{Optimization for truss design using Bayesian optimization}

\markboth{}{June 2023\hb}

\author{\IEEEauthorblockN{Bhawani Sandeep\IEEEauthorrefmark{1},
Surjeet Singh\IEEEauthorrefmark{2}, and Sumit Kumar\IEEEauthorrefmark{3}}\\ 
\IEEEauthorblockA{\IEEEauthorrefmark{1}\IEEEauthorrefmark{2} KL University, Guntur, India,
\IEEEauthorrefmark{3}Georgia State University,USA\\
Email: Corresponding author: skumar18@student.gsu.edu}}

\maketitle

\begin{abstract}
 In this work, geometry optimization of mechanical truss using computer-aided finite element analysis is presented. The shape of the truss is a dominant factor in determining the capacity of load it can bear. At a given parameter space, our goal is to find the parameters of a hull that maximize the load-bearing capacity and also don't yield to the induced stress. We rely on finite element analysis, which is a computationally costly design analysis tool for design evaluation.
 For such expensive to-evaluate functions, we chose Bayesian optimization as our optimization framework which has empirically proven sample efficient than other simulation-based optimization methods. 
 By utilizing Bayesian optimization algorithms, the truss design involves iteratively evaluating a set of candidate truss designs and updating a probabilistic model of the design space based on the results. The model is used to predict the performance of each candidate design, and the next candidate design is selected based on the prediction and an acquisition function that balances exploration and exploitation of the design space.
 Our result can be used as a baseline for future study on AI-based optimization in expensive engineering domains especially in finite element Analysis. 
\end{abstract}


\input{1.Introduction.tex}
\input{3.Approach.tex}

\input{4.Experiments.tex}

\input{2.RelatedWorks.tex}

\input{6.ConclusionFutureWork.tex}

\bibliographystyle{IEEEtran}
\bibliography{references}
\end{document}

%% file: 1.Introduction.tex
\section{Introduction and Problem Formulation}
\label{sec:introduction}

A truss is a structural element that is commonly used in engineering and architecture to support loads over long spans. It is a system of interconnected triangles that work together to distribute loads in a manner that is both efficient and stable.
Trusses are commonly used in a variety of applications, such as bridges, roofs, towers, and even bicycles. In each case, the truss is designed to meet specific requirements for load capacity, span length, and environmental conditions.
For example, in bridge construction, trusses are often used to support the weight of the deck and distribute it evenly to the piers or abutments. By using a truss system, the bridge can span long distances without requiring additional support structures in the middle. This can be particularly useful in areas where there are obstacles such as waterways, valleys, or other natural features. The use of mechanical trusses is becoming increasingly popular in a variety of applications, such as bridge construction, roof structures, and even the design of complex machinery. 
Another advantage of trusses is their flexibility. They can be designed to be lightweight and easy to transport, making them ideal for use in temporary structures or in remote locations where transportation and logistics are a concern.
In roof construction, trusses are used to support the weight of the roof and distribute it evenly to the walls or columns. By using a truss system, the roof can span large distances without requiring additional support structures in the middle. This can be particularly useful in areas where there is a need for large open spaces, such as sports arenas or convention centers.
In tower construction, trusses are often used to provide stability and support to tall structures. By using a truss system, the tower can resist the forces of wind and other environmental factors that can cause it to sway or tip over.
In bicycle construction, trusses are used to provide a lightweight and strong frame that can withstand the forces of pedaling and other stresses. By using a truss system, the bike can be made lighter and more efficient without sacrificing strength or durability.

Overall, trusses are a versatile and effective structural element that is widely used in engineering and architecture. Whether in bridges, roofs, towers, or bicycles, trusses provide a strong and stable platform that can support large loads over long spans. With their combination of efficiency, strength, and aesthetic appeal, trusses are sure to remain a popular choice for designers and engineers for many years to come. One area where mechanical trusses have shown great potential in recent times is in the field of renewable energy. They are being used to support large-scale solar arrays or wind turbines, allowing for optimal positioning and orientation of the equipment to maximize energy production. The ability to adjust the trusses allows for fine-tuning of the system to ensure that it is operating at maximum efficiency, even in changing weather conditions.

The truss design problem is highly nonlinear and its convexity is not guaranteed on a given design space. 
Truss optimization is the process of designing trusses that can carry the required loads while minimizing the weight and resultantly cost of the structure. There are several methods for optimizing trusses, including manual optimization, computer-aided optimization, and genetic algorithms.
Manual optimization involves the use of mathematical equations and hand calculations to determine the optimal size and spacing of truss members. This method can be time-consuming and requires a high level of expertise, but it can be effective for simple truss structures.
Computer-aided optimization involves the use of software programs that use Finite Element Analysis (FEA) to analyze and optimize truss designs. These programs can quickly analyze a large number of design options and provide recommendations for the most efficient and cost-effective truss design. Computer-aided optimization can be used for more complex truss designs, but it requires specialized software and expertise to use effectively.

In this work, we focus on computer-aided optimization of truss geometric design using the AI-based optimization framework Bayesian optimization. Application of Bayesian optimization in engineering design is coming up in various recent works\cite{vardhan2023search,vardhan2023constrained}. Bayesian optimization has empirically shown sample efficiency in expensive-to-evaluate processes. Since FEA is an expensive process\cite{vardhan2022data}, we leverage the BO in this scenario. We observed that within $100$ evaluations, BO is able to find near-optimal design. This study can be used as a baseline for optimization studies in  expensive to-evaluate engineering domains.  

%% file: 3.Approach.tex
\section{Approach}
\label{sec:approach}
\subsection{Bayesian optimization:}
Bayesian optimization\cite{frazier2018tutorial} is a sequential model-based optimization technique that is used to optimize black-box functions, where the objective function is unknown and expensive to evaluate. It works by iteratively constructing a probabilistic model of the objective function, based on the data collected from the previous iterations, and using the model to decide the next point to evaluate.
In Bayesian optimization, the objective function is assumed to be a sample from a Gaussian process ($\mathcal{GP}$)\cite{rasmussen2003gaussian}, which is a collection of random variables, any finite number of which have a joint Gaussian distribution. The $\mathcal{GP}$ is defined by its mean and covariance functions, which capture our prior belief as well as data-based behavior of the objective function.
At each iteration, the GP model is updated with the new data, and a so-called acquisition function is used to decide the next point to evaluate. The acquisition function balances the exploration of new regions of the search space with the exploitation of regions that are likely to contain the optimum. Common acquisition functions include Expected Improvement (EI)\cite{jones1998efficient}, Probability of Improvement (PI), and Upper Confidence Bound (UCB)\cite{srinivas2012information}.
The iterative process continues until a stopping criterion is met, such as a maximum number of iterations or a small improvement in the objective function. The final solution is then taken as the point that maximizes the objective function.
Bayesian optimization has been successfully applied in various domains, such as hyperparameter tuning in machine learning, experimental design in materials science, and drug discovery in bioinformatics. It is particularly useful when the evaluation of the objective function is expensive or time-consuming, as it can efficiently explore the search space and find the optimum with a small number of evaluations.
Sure! The mathematical equations for Bayesian optimization are as follows:

Let $f(x)$ be the unknown black-box function to be optimized, where $x \in X$ is a d-dimensional input vector from a bounded domain $X$.

A Gaussian process ($\mathcal{GP}$) prior over $f(x)$ is assumed:
$$f(x) \sim \mathcal{GP}(m(x), k(x, x'))$$
where $m(x)$ is the mean function, and $k(x, x')$ is the covariance function.
At any iteration $t$, a set of observations $D_t = {(x_0, y_0),...(x_i,y_i),...,(x_t,y_t)}$ is collected, where $y_i = f(x_i) + \epsilon_i$ is the noisy observation of $f(x_i)$, and $\epsilon_i$ is the measurement noise.
Based on the observations $D_t$, the $\mathcal{GP}$ posterior over $f(x)$ is updated using Bayes' rule:
$$p(f(x)|D_t) = N(f(x)|m_t(x), k_t(x, x'))$$
where $m_t(x)$ and $k_t(x, x')$ are the posterior mean and covariance functions at $t^{th}$ iteration respectively.
An acquisition function $\alpha(x)$ is defined to choose the next point $x_{t+1}$ to evaluate. A common acquisition function is Expected Improvement (EI):
$$EI(x) = E[max(y^* - y, 0)]$$
where $y^*$ is the best observation so far, $y^* = max(y_1, ..., y_t)$, and $\mathbf{E}[.]$ denotes the expectation with respect to the posterior distribution $p(f(x)|D_t)$.

Finally, the next point to evaluate $x_{t+1}$ is chosen as:
$$x_{t+1} = argmax_x \alpha(x)$$
The process continues until a stopping criterion is met, such as a maximum number of iterations or a small improvement in the objective function.
The specific form of the mean and covariance functions, as well as the choice of acquisition function, is a hyperparameter setting but can be varied can vary depending on the user's preference. 

\subsection{Finite element analysis and truss design:}Finite element analysis (FEA) is a numerical technique used to analyze the behavior of complex structures. FEA involves dividing the structure into a large number of smaller, interconnected elements, each of which is modeled using mathematical equations that describe its behavior. The equations for each element are then combined to create a system of equations that describe the behavior of the entire structure. This system of equations can be solved using numerical methods to obtain the stresses and strains in each element.

In truss design, FEA is used to determine the stresses and strains in the truss members to determine the location of joints and angle of the truss that can withstand the loading condition without failure of the design. In spite of using any commercial tool such as ANSYS, Abaqus, or SolidWorks, we used freeCAD\cite{riegel2016freecad} as our simulation tool. Under given loading conditions, we run simulations to obtain the stress and strain distributions and try to minimize the weight of the truss within the given range of free parameters.  For loading conditions and related failure, we use von-mises stress obtained from the FEA simulation and compare it with the yield strength of the material along with some safety factors\cite{vardhan2022reduced}. 

%% file: 4.Experiments.tex
\section{Experiment and its result}

\label{sec:experiments}
\begin{figure}[h!]
        \centering
        \includegraphics[width=0.95\textwidth]{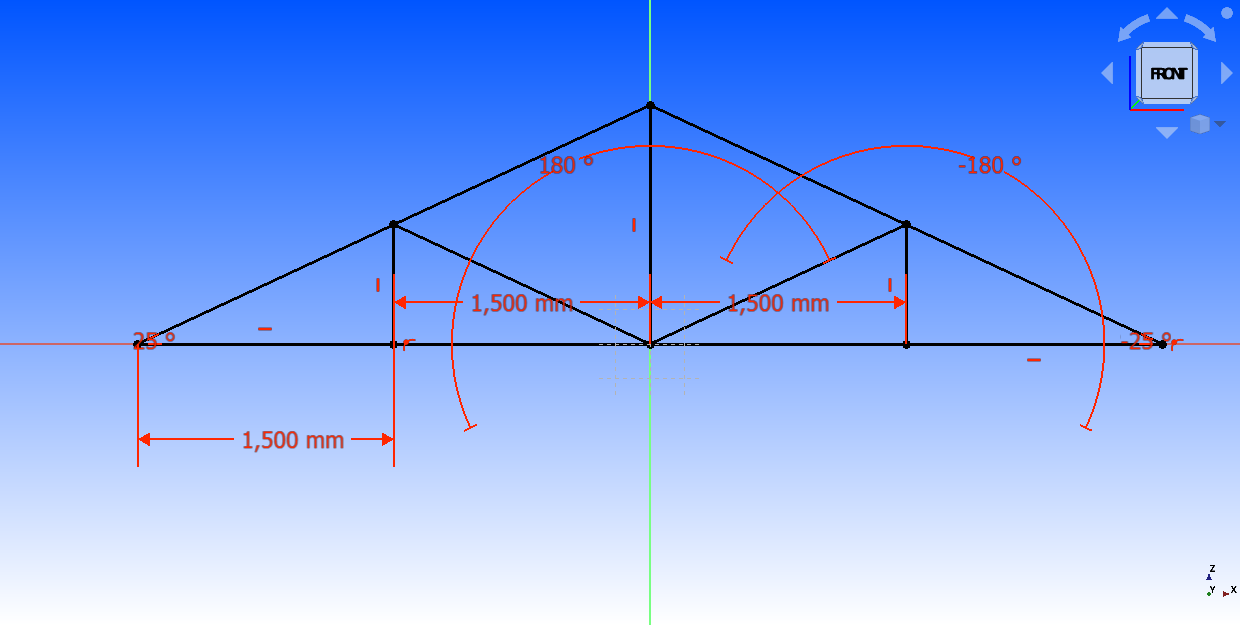}
        \caption{Truss sketch, its parameter of variations}
        \label{fig:truss_sketch}
\end{figure} 
Figure \ref{fig:truss_sketch} shows the sketch of the truss. Its design is split into 6 triangles organized and constrained at joints. The parameter that can be varied $a$,$b$,$c$, $\theta_1$ and $\theta_2$. The range and the definition is shown in table \ref{tab:ds}. 
Since our focus of the design is the truss used in outdoor applications where interaction with water and moisture and other factor is common along with being light-weighted for easy implementation, we choose AL-6061T6 as our material. Aluminum trusses are lightweight, strong, and resistant to corrosion, making them ideal for use in outdoor structures and portable stages. The specific property of this material is density $2700 kg/m^3$, Young's modulus $70.0 GPa$, and Poisson ratio $0.350$. 

\begin{figure}[h!]
        \centering
        \includegraphics[width=0.95\textwidth]{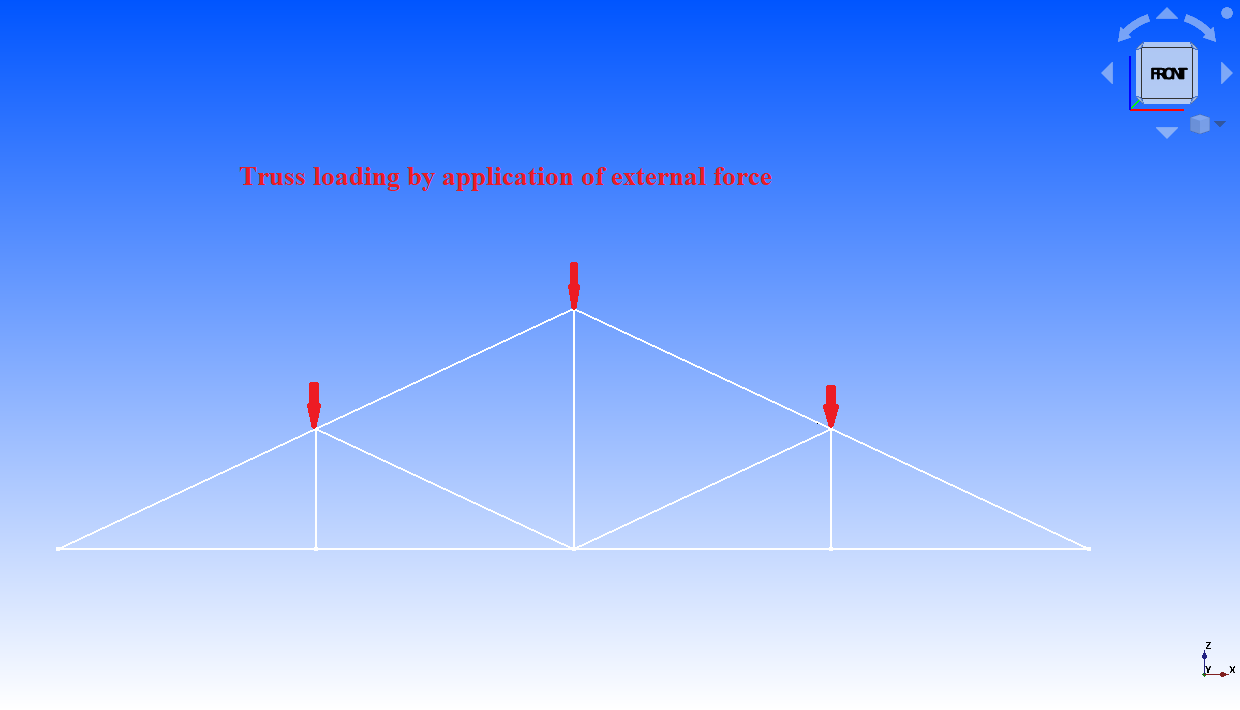}
        \caption{Truss and point and direction of applied force. The total force applied is 12,000 Newton, assuming the maximum loading capacity of the truss is 1000 Kg and a safety factor of $1.2$.}
        \label{fig:truss_loading}
\end{figure}

\begin{table}
\centering
\normalsize
\begin{tabular}{cccc} 
\hline
\hfil \textbf{Parameter} & \hfil \textbf{Symbol} &   \hfil \textbf{ Minimum } & \hfil \textbf{ Maximum } \\
\hline
 Left outer section & $a$ & $500$  & $2500$ mm \\
left center section & $b$ & $500$ & $2500$ mm \\
Right center section & $c$ & $500$ & $2500.0$ \\
Right side angle & $\theta_1$ & $0^o$ & $60^o$  \\
Left side angle & $\theta_2$ & $0^o$ & $60^o$  \\
Right outer section $(d)=8000-(a+b+c)$ &  &  \\
\hline
\end{tabular}
\vspace{0mm}
\caption{Range of design parameters for optimization}
\label{tab:ds}
\end{table}
The applied force on the truss is distributed on three loading points as shown in figure \ref{fig:truss_loading}. We expect our truss to work under $1000 Kg$ loading conditions, and we also keep a safety factor of $1.2$. Accordingly, the total applied force on the truss joint is $12000$ Newtons (refer figure \ref{fig:truss_loading}). 

Since the evaluation of FEA is computationally intensive, we keep our optimization budget for $100$ evaluations. The final parameter that we found  after optimization is the optimal design parameters: $a= 1200.0$; $b= 2497.3$; $c= 2498.2$; $d= 1804.5$;$\theta_1= 42.0$; $\theta_2= 45.0$. For running Bayesian optimization, we use the GPyopt\cite{gpyopt2016} python package and integrate it with the FEA evaluation tool FreeCAD. These experiments empirically show that the AI has potential to revolutionize the field of design and other domains \cite{vardhan2021rare}. 

%% file: 2.RelatedWorks.tex
\section{Related Work}
\label{sec:related_works}

Initial work on an automatic truss design problem, formulated based on ground structure approach by \cite{dorn1964automatic}, who formulated it as a set of joints distributed in the design space and connected by bars. 
Later this is extensively studied by many researchers, like  Dobbs and Felton\cite{dobbs1969optimization}, Bendsøe et al. \cite{bendsoe1994optimization}, Miguel and Miguel \cite{miguel2012shape} etc. 
Due to the high nonlinearity of the problem, several solutions have been proposed, which aim to improve the computation tractability by simplification of problems, e.g., Imai et al\cite{imai1981configuration}, Ringerts \cite{ringertz1985topology}, Achtziger\cite{achtziger2007simultaneous}, and He et al \cite{he2015rationalization}. The most commonly used technique is called the alternating method where the problems are solved for a fixed geometry but with a variation in topology and vice versa in a block-coordinate manner to obtain designs that satisfy the optimality conditions for the  subproblems. 
Another group of studies on the optimization of trusses attempts to incorporate stronger constraints than the general formulations to improve the practicality of the optimal designs. These constraints are applied to stresses, local buckling, global stability constraints, beam modeling, etc and found in works of Kirsch\cite{kirsch1990singular}, Stolpe et al \cite{stolpe2001trajectories}, Rozvany \cite{rozvany1996difficulties}, Guo et al. \cite{guo2005optimum}, Descamps et al \cite{descamps2014nominal}), Ben-Tal et al. \cite{ben2000optimal}, Stingl \cite{stingl2006solution} and Tugilimana et al. \cite{tugilimana2018including}, Torii et al.\cite{torii2015modeling} etc.
Another approach to truss design problems is by  formulating it as a nonlinear semidefinite programming problem that involves assigning global stability constraints using a linear buckling model and extensively studied by Kanno et al.\cite{kanno2001sequential}, Levy and Su \cite{levy2004modeling}, Kočvara \cite{kovcvara2002modelling}, Stingl \cite{stingl2006solution}, and Evgrafov \cite{evgrafov2005globally}, Kočvara and Stingl\cite{evgrafov2005globally} etc. On the optimization front, apart from the active learning approach \cite{vardhan2022deepal} various traditional approaches have been developed in the past. 
A customized  primal-dual interior point method is deployed that exploits the sparsity and low-rank properties of the associated element stiffness matrices  by Ben \cite{ben1993bendsoe}, Achtziger et al.\cite{achtziger1992equivalent}; Gilbert et al\cite{gilbert2003layout}, Weldeyesus et al\cite{weldeyesus2018specialized} etc. Application of machine learning in engineering design is shown in \cite{vardhan2021machine,li2023construction,vardhan2022deep}, where learning models are used to explore and learn the design space and then by capitalizing the generalization capability able to find designs on new requirements or noble design.  Few new hybrid approaches are also being developed in recent works\cite{vardhan2023fusion}.

%% file: 6.ConclusionFutureWork.tex
\section{Conclusion and Continuation}
\label{sec:conclusionFutureWork}

In this work, we presented the conclusions of our experiment on truss design. The experiment aimed to design optimization and find the optimal parameter of the truss under a given requirement. 
Through the use of finite element analysis (FEA), we analyzed the stress and deformation levels in each truss configuration under given loading conditions. We used Bayesian optimization with EI as our acquisition function to optimize our design. 
Nonetheless, we recommend that further studies are conducted on truss design, especially on other loading conditions and different materials, to provide more comprehensive results that can inform the optimal design of trusses for a wide range of applications.